\documentclass[pra,floatfix,footinbib,reprint,superscriptaddress]{revtex4-2}
\usepackage{ytableau}
\usepackage{graphicx}
\usepackage{dcolumn}
\usepackage{times}
\usepackage{caption}
\usepackage{physics}
\usepackage{amsmath, amssymb, dsfont}
\usepackage{bbold}
\usepackage{amsthm}
\usepackage{amsfonts}
\usepackage{tabularx}
\usepackage{subfigure}
\usepackage{tikz}
\usepackage{xcolor}
\usepackage{tcolorbox}
\usepackage{natbib}
\usepackage{graphicx}
\usetikzlibrary{matrix}
\usepackage[pdftex, colorlinks, allcolors=blue]{hyperref}
\usepackage[T1]{fontenc}
\usepackage[utf8]{inputenc}

\begin{document}

\title{A One-Dimensional Electron Gas Coupled to Light}

\author{Victor Bradley}
\affiliation{Texas Tech University, Department of Physics and Astronomy, Lubbock, Texas 79409, USA}

\author{Kamal Sharma}
\affiliation{Texas Tech University, Department of Physics and Astronomy, Lubbock, Texas 79409, USA}

\author{Mohammad Hafezi}
\affiliation{Joint Quantum Institute,
NIST/University of Maryland, College Park, Maryland,
20742, USA}
\affiliation{Institute for the Research in
Electronics and Applied Physics, University of Maryland,
College Park, Maryland 20742, USA}

\author{Wade DeGottardi}
\affiliation{Texas Tech University, Department of Physics and Astronomy, Lubbock, Texas 79409, USA}

\date{\today}

\begin{abstract}
The prospect of using light to probe or manipulate quantum materials has become an active area of interest. Here, we investigate a quantum wire---treated as a finite-sized one-dimensional electron gas---that is coupled to a single photonic mode. This work focuses on the radiative properties of the wire when it is prepared in some excited state. One of our key results addresses the photon cascade initiated by the creation of a single electron-hole pair. Repeated photon emission leads to the generation of entanglement between the electron and hole and the emission rates exhibit Dicke-like superradiance. In general, the radiation is characterized by a competition between superradiance and Pauli blocking. We find that the one-dimensional electron gas represents an ideal system for investigating quantum coherent phenomena and quantum entanglement. This work has direct relevance to applications in quantum computation and quantum transduction.
\end{abstract}
\maketitle

\section{Introduction} 

Light-matter coupling underlies many of the experimental probes of condensed matter systems. Typical experiments are conducted in the linear response regime~\cite{drake_springer_2006} for which a system excited by an external perturbation relaxes before it can be excited again. On the other hand, quantum coherent effects arise when a probe interacts with a system many times before it decoheres. Thus, the quantum regime requires strong coupling between the system and probe~\cite{bloch_strongly_2022,rivera_lightmatter_2020,roux_strongly_2020,frisk_kockum_ultrastrong_2019}. The inherent weakness of light-matter coupling---dictated by the fine structure constant $\alpha = 1/137$---would seem to preclude accessing the quantum regime using light. However, the strong coupling regime can be achieved using cavity QED~\cite{bloch_strongly_2022,rivera_lightmatter_2020,roux_strongly_2020,schlawin_cavity_2022,vasanelli_ultra-strong_2016, kroeger_continuous_2020,frisk_kockum_ultrastrong_2019}. The possibility of probing or even altering quantum systems using cavity QED is now receiving a great deal of attention~\cite{bloch_strongly_2022,rivera_lightmatter_2020,roux_strongly_2020,baskaran_superradiant_2012, vasanelli_ultra-strong_2016, eckhardt_quantum_2022, rokaj_free_2022, schlawin_cavity_2022,manasi_light_2018,rivera_lightmatter_2020,
yang_lightwave-driven_2019,mootz_visualization_2022,kovalev_proposal_2020,luo_quantum_2022, mivehvar_cavity_2021, baumann_dicke_2010, frisk_kockum_ultrastrong_2019, kroeger_continuous_2020}. Light-coupled many-body systems have been proposed as a way of realizing quantum computation~\cite{ayral_quantum_2023} and quantum transduction~\cite{lauk_perspectives_2020}. More ambitious proposals call for the use of light to fundamentally alter the properties of condensed matter systems~\cite{bloch_strongly_2022}.

While this interest has been primarily stimulated by recent advances in cavity QED technology~\cite{bloch_strongly_2022}, the study of light-coupled matter in the quantum coherent regime has a long history. The Dicke model, introduced nearly seventy years ago~\cite{dicke_coherence_1954}, describes $N$ non-interacting spin-1/2 degrees of freedom coupled to a photonic mode~\cite{garraway_dicke_2011, gross_superradiance_1982, vertogen_dicke_1974,kirton_introduction_2019, garraway_dicke_2011}. The assumption that the spins retain their phase coherence leads to the Dicke model's signal prediction of \emph{superradiance}. This phenomenon describes the increase in the rate of photon emission of a system due to the quantum entanglement that this emission itself generates~\cite{gross_superradiance_1982}.

\begin{figure}
\begin{center}
\includegraphics[height = 5cm]{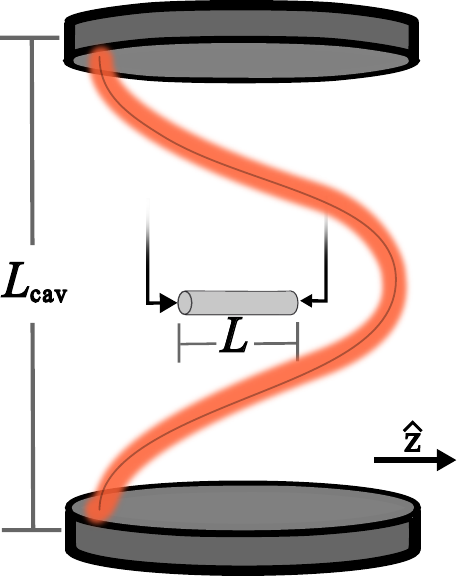}
\caption{A quantum wire in a cavity coupled to a single photonic mode. Tunneling leads attached to the end of the wire, indicated by arrows, allow for the preparation of excited states through the injection of electrons and holes into the system. These leads can also be used to perform conductance measurements.}
\label{fig:QW}
\end{center}
\end{figure}

In this work, we study a one-dimensional electron gas coupled to light in the quantum coherent regime. There has been considerable work investigating low-dimensional quantum gases and liquids coupled to light~\cite{vasanelli_ultra-strong_2016,mivehvar_cavity_2021,rokaj_free_2022,eckhardt_quantum_2022,manasi_light_2018,pivovarov_quantum_2020}. Investigations of Dicke superradiance in condensed matter systems have naturally tended to focus on systems in which spin or pseudo-spin degrees of freedom decay~\cite{baskaran_superradiant_2012}. Here, we explore the emergence of superradiance in a different context: a spin-polarized finite-sized one-dimensional electron gas. A possible experimental realization, shown in Fig.~\ref{fig:QW}, involves a quantum wire coupled to a lossy photonic cavity. The cavity plays several key roles. Most importantly, it enhances light-matter coupling, allowing the electron gas to emit or absorb multiple photons before quantum coherence is lost. The discreteness of the cavity's spectrum allows for coupling to electronic transitions of a specific frequency. The antenna-like nature of the quantum wire is ideal for selectively coupling to specific cavity modes~\cite{joh_single-walled_2011}. A carbon nanotube is a good candidate for the quantum wire. Nanotubes have attracted considerable interest in the context of cavity QED~\cite{mergenthaler_circuit_2021}.

For a variety of excited states, we identify the intermediate states of the system as it relaxes. At each step in this cascade, photon emission rates are calculated. A paradigmatic example is the cascade initiated by an electron-hole pair, which represents a fundamental excitation of Fermi gases. We find that the repeated emission of photons generates momentum-space entanglement~\cite{flynn_momentum_2023} between the electron and hole, which in turn leads to Dicke-like superradiance. The rate of photon emission continues to increase until either the electron or hole reaches the Fermi surface. The example of two excited electrons is considerably richer because the higher energy electron can be Pauli blocked by the lower energy one. This frustration affects the photon emission rates and leads to a complex many-body state.

The paper is organized as follows. In Sec.~\ref{sec:spontaneous}, we introduce the physical setup of interest: a quantum wire coupled to a photonic mode. In Sec.~\ref{sec:bosonization}, the description of the 1D electron gas using bosonization is given and its application to light-matter coupling is illustrated. Sec.~\ref{sec:cascades} presents several paradigmatic cascades. In Sec.~\ref{sec:band_curvature}, we consider the effects of the curvature of the electron dispersion as well as electron-electron interactions. Finally, in Sec.~\ref{sec:conclusion}, we present our conclusions and outlook for future work.

\section{Physical Setup}

\label{sec:spontaneous}

The physical setup, shown in Fig.~\ref{fig:QW}, consists of a quantum wire of finite length $L$ in a photonic cavity. The electrons in the wire are modeled as a non-interacting 1D electron gas, which for theoretical convenience is taken to be spin-polarized. The electronic Hamiltonian is
\begin{equation}
H_0 = \sum_{j=-\mathcal{N}+1}^\Lambda \hbar \omega_j c_j^\dagger c_j^{\phantom\dagger},
\label{eq:H0}
\end{equation}
where the operator $c_j^\dagger$ creates an electron in the $j^{\textrm{th}}$ single-electron state, which has an energy $\hbar \omega_j$. The integer $j$ indexes the single-particle states, as shown in Fig.~\ref{fig:states}a. We take $j = 0$ to be the highest energy single-particle state that is occupied when the electron gas is in the many-body ground state $|G\rangle$. The number of electrons in the wire is $\mathcal{N}$ and the integer $\Lambda \gg 1$ is a high energy cut-off. We focus on the case of a linear dispersion $\omega_j = v_F j / L$ where $v_F$ is the Fermi velocity. The effects associated with a massive dispersion will be addressed in Sec.~\ref{sec:band_curvature}. 

The quantum wire is coupled resonantly with one of the cavity modes. The frequency $n \omega_1 = \pi n v_F/L$ of the gas, where $n$ is an integer, is matched to the frequency of (say) the fundamental cavity mode, $\pi c / L_{cav}$, where $c$ is the speed of light and $L_{cav}$ is a characteristic length of the cavity. The resonance condition gives $L/L_{cav} = n v_F / c$. As long as $n v_F \ll c$, we have $L \ll L_{cav}$. Since the wavelength of light $\lambda \sim L_{cav}$, the condition $L \ll L_{cav}$ ensures that $L \ll \lambda$ and so the dipole approximation applies~\cite{scully_quantum_1997}. The coupling of the cavity mode to the wire is
\begin{equation}
H_{c} = - E_z \wp,
\label{eq:e-c}
\end{equation}
where $E_z$ is the component of the electric field along the wire (taken to be the $z$-direction) and $\wp$ is the dipole operator of the electron gas. The dipole operator is $\wp = e \int dz \, z \, c_z^\dagger c_z^{\phantom\dagger}$, where $c_z$ annihilates an electron at $z$ and $e$ is the electron charge. The operator $c_z$ can be written as $c_z = \sum_j \int dz \, \psi_j (z) c_j$, where $\psi_j(z)$ is the single-particle functions for the $j^{th}$ single-electron state. Substituting this mode expansion into the expression for the dipole operator gives $\wp = \sum_{jk} \mathcal{M}_{jk} c^\dagger_j c^{\phantom\dagger}_k$, where
\begin{equation}
\mathcal{M}_{jk} = \int_{-L/2}^{L/2} dz \, \psi_{j}^\ast(z) \psi_k^{}(z) z.
\end{equation}
The amplitudes $\mathcal{M}_{jk}$ are simply the dipole matrix elements between the first-quantized wave functions. 

We work in the so-called `bad cavity' regime in which the photon emission rate from the quantum wire is much less than the cavity photon loss rate. In this limit, the rate of photon emission from the wire is given by Fermi's golden rule. The decay rate of the state $|\alpha\rangle$ via the emission of a photon with frequency $n\omega_1$ is~\cite{gross_superradiance_1982} 
\begin{equation}
\Gamma_{\alpha} = \gamma_n  \langle \alpha | \mathcal{D}_n^\dagger \mathcal{D}_n | \alpha \rangle, 
\label{eq:gamma_n}
\end{equation}
where
\begin{equation}
\mathcal{D}_n = \sum_{m} c_{m-n}^\dagger c_{m}^{\phantom\dagger}.
\label{eq:bigd}
\end{equation}
The rate $\gamma_n$ depends on $\mathcal{M}_{jk}$ and is thus sensitive to the nature of the single-particle wave functions $\psi_j(z)$. In Appendix A, we consider a particular form of the $\psi_j(z)$ for which
\begin{equation}
\gamma_n = \frac{2 e^2 L^2 E_z^2}{\pi^3 n^4 \hbar }  G(n \omega_1)
\label{eq:gamma_n_def}
\end{equation}
for $n$ odd, while it vanishes for $n$ even. The quantity $G(n\omega_1)$ is the density of states of the cavity modes at the frequency $n \omega_1$. 

The operator $\mathcal{D}_n$ plays the role of a many-body lowering operator, taking $| \alpha \rangle$ to $| \alpha' \rangle$, its daughter state after the decay. We have
\begin{equation}
| \alpha' \rangle = \frac{\mathcal{D}_n | \alpha \rangle}{\sqrt{ \langle \alpha | \mathcal{D}_n^\dagger \mathcal{D}_n | \alpha \rangle}}.
\label{eq:demotion}
\end{equation}
Tacit to Eq.~(\ref{eq:demotion}) is that the matrix element $\mathcal{M}_{jk}$ is a function of the difference $j-k$, but is only weakly dependent on $j+k$. This assumption is discussed in Sec.~\ref{sec:band_curvature}.  

\section{Bosonization}

\label{sec:bosonization}

The description of the electron gas in terms of electronic excitations can quickly become unwieldy because the operator $\mathcal{D}_n$ tends to generate complicated superpositions. Fortunately, the bosonization technique offers an alternative basis~\cite{stone_bosonization_1994} that is particularly useful in the present context. As we will see, it also can elucidate physics that is obscure in terms of the electrons. The bosonization technique provides a one-to-one mapping between the excitations of a 1D electron system and a gas of non-interacting bosons~\cite{stone_bosonization_1994}. The Hamiltonian for these bosons is
\begin{equation}
H_B = \hbar\omega_1 \sum_{k=1}^\infty k \,  b_k^\dagger b_k^{\phantom\dagger},
\label{eq:HB}
\end{equation}
where $b_k$ destroys a boson in the $k^{\textrm{th}}$-mode. This Hamiltonian describes an infinite number of harmonic oscillators with frequencies $\omega_1$, $2 \omega_1$, $3 \omega_1$, etc. The eigenstates of $H_B$ are Fock states $| l_1 l_2 ... \rangle_b$, where $l_i$ is the occupation number of the $i^{\textrm{th}}$ bosonic mode. According to Eq.~(\ref{eq:HB}), the total energy of this state is $E =  l_1 \hbar \omega_1 + l_2 \left( 2 \hbar \omega_1 \right) + ...$

\begin{figure}
\begin{center}
\includegraphics[width = \linewidth]{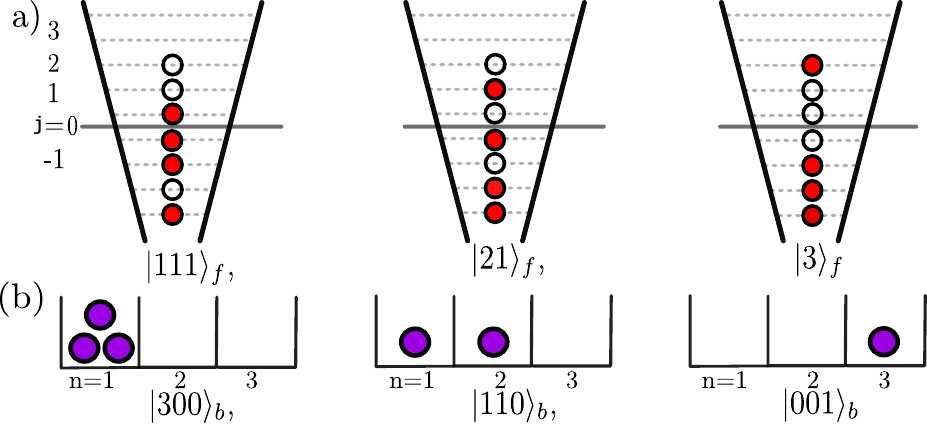}
\caption{Three excited states of the electron gas with energy $3 \hbar \omega_1$ in the (a) electronic (fermionic) and (b) bosonic bases.}
\label{fig:states}
\end{center}
\end{figure}

To motivate this approach, consider the operator $\mathcal{D}_1^\dagger\mathcal{D}_1^{\phantom\dagger}$ that appears in the expression for $\Gamma_n$ in Eq.~(\ref{eq:gamma_n}). We consider this operator in the subspace of the three excited states with energy $3\hbar \omega_1$. The degeneracy of this subspace is equal to the partitions of the integer $3$, i.e., the number of distinct ways $3$ can be written as a sum of integers: $3$, $2+1$, or $1+1+1$~\cite{stone_bosonization_1994}. These partitions correspond to the states $|3\rangle_f$, $| 2 1 \rangle_f$, and $|111\rangle_f$, respectively, where $| \lambda_1 \lambda_2 \lambda_3 ... \rangle_f$ is the excited state formed from the ground state by promoting the most energetic electron by an energy $\lambda_1 \hbar \omega_1$, the second-most energetic electron by $\lambda_2 \hbar \omega_1$, and so on. Pauli exclusion requires that these integers satisfy $\lambda_1 \geq \lambda_2 \geq \lambda_3...$ For instance, the state $|3 \rangle_f = c_3^\dagger c_{0}^{\phantom\dagger} |G \rangle$. These three states are depicted in Fig.~\ref{fig:states}a. The matrix elements of $\mathcal{D}_1^\dagger\mathcal{D}_1^{\phantom\dagger}$ are  
\begin{equation} 
\mathcal{D}_1^\dagger \mathcal{D}_1 = \left(
  \begin{array}{ccc}
    1 & 1 & 0 \\
    1 & 2 & 1 \\
    0 & 1 & 1 \\
  \end{array}
\right),
\label{eq:matrix}
\end{equation}
in the basis that $|3 \rangle_f, |2 1 \rangle_f,$ and $|111\rangle_f$ are represented by $(1 \, 0 \, 0)^T$, $(0 \, 1 \, 0)^T$, and $(0 \, 0 \,1)^T$, respectively.

The eigenvalues of Eq.~(\ref{eq:matrix}) are $\{ 3, 1, 0 \}$. The significance of these numbers becomes apparent in the bosonic basis. Naturally, there are three bosonic states with energy $3 \hbar \omega_1$, namely $|300\rangle_b =  \left(b_1^\dagger\right)^3 | G \rangle$, $|1 1 0 \rangle_b = b_1^\dagger b_2^\dagger|G \rangle$, and $|001\rangle_b = b_3^\dagger |G \rangle$. Apparently, the eigenvalues $3$, $1$, and $0$, correspond to the occupation number $l_1$ of these three states. The operator $\mathcal{D}_1^\dagger \mathcal{D}_1^{\phantom\dagger}$ is thus diagonal in the the bosonic basis. This is true in general and follows from the  bosonization identity~\cite{stone_bosonization_1994}
\begin{equation}
b_n = \frac{1}{\sqrt{n}} \mathcal{D}_n.
\label{eq:bosid}
\end{equation}
For $n = 1$, the operator $\mathcal{D}_1^\dagger \mathcal{D}_1 = b_1^\dagger b_1^{\phantom\dagger}$, which is simply the number operator for the $k=1$ mode. There is not a simple correspondence between the states in the fermionic and bosonic bases. For example, the first state in Fig.~\ref{fig:states}a is 
\begin{equation}
|3 \rangle_f = \frac{1}{\sqrt{6}} | 300 \rangle_b + \frac{1}{\sqrt{2}} |210 \rangle_b + \frac{1}{\sqrt{3}} | 001 \rangle_b. 
\label{eq:3f}
\end{equation}
A general expression for the unitary transformation exists and is given by Eq.~(\ref{eq:U}) in Appendix~\ref{sec:unitary}. It was first established in Ref.~\cite{stone_bosonization_1994}.

\section{Cascades}

\label{sec:cascades}

This Section presents our key results. We consider the sequence of many-body states of the gas as photons are emitted,
\begin{equation}
||1\rangle \! \rangle \xrightarrow{\Gamma_1} ||2\rangle \!  \rangle \xrightarrow{\Gamma_2 }
 \ldots \xrightarrow{\Gamma_{N-1}} || N \rangle \! \rangle,
\label{eq:cascade}
\end{equation}
i.e., $||1 \rangle \! \rangle$ is the initial (electronic) state in the cascade and $|| N \rangle \! \rangle$ is the last. The decay rate of the state $|| m \rangle \! \rangle$ is denoted by $\Gamma_m$. (The double brace notation indicates that the state is labeled by the order it appears in the cascade.) 

The states in the cascade can be obtained by applying the operator $\mathcal{D}_n$ to the initial state $|| 1 \rangle \! \rangle$ repeatedly. The subsequent states and their decay rates are given by Eqs.~(\ref{eq:demotion}), (\ref{eq:bosid}) and (\ref{eq:gamma_n}). For example, for the case $n = 1$ (for which the cavity mode is resonant with the electronic transition frequency $\omega_1$), the various states in the cascade (\ref{eq:cascade}) can be obtained by repeatedly applying the bosonic lowering operator $b_1$. The cascade terminates with the state $|| N \rangle \! \rangle$ that is annihilated by $b_1$, i.e., $ b_1 || N \rangle \! \rangle = 0$. In certain cases, the bosonic basis is unwieldy and it is advantageous to work in the fermionic basis.

We work in the regime in which the electron gas retains phase coherence throughout the cascade. The expected time for the cascade to occur is $\sum_m 1/\Gamma_m$. This time must be less than the lifetime $\tau$ of the gas due to non-radiative transitions. On the other hand, we assume that $\gamma_1 \ll \omega_1$ so that electron-light coupling is weak and the electronic states and cavity mode do not strongly hybridize. Strong light-matter coupling is necessary to achieve this intermediate regime.

\subsection{Excited Bosonic Mode}

We consider $N$ bosonic excitations in the $k=1$ mode, i.e., $||1 \rangle  \! \rangle = |N,000...\rangle_b$. This state can be prepared by letting a quantum wire in its ground state absorb $N$ photons with frequencies $\omega_1$. The intermediate states in the cascade, obtained by repeatedly applying $b_1$ to the initial state, are
\begin{equation}
|| m \rangle \! \rangle = |N-m+1,000...\rangle_b.
\end{equation}
The decay rate of these states is given by Eq.~(\ref{eq:gamma_n}),
\begin{equation}
\Gamma_m = \gamma_1 (N-m+1).
\label{eq:rates1}
\end{equation}
This result has a particularly simple interpretation: the rate reflects the fact that each bosonic excitation can decay independently and thus $\Gamma_m$ is proportional to the number of bosons remaining in the $k = 1$ mode. 

This example reveals a simple but profound aspect of the 1D electron gas with a linear spectrum---it exhibits trivial fluorescence. If the gas absorbs some number of photons of various energies, then it can only relax by emitting the exact same population of photons. Although this fact is not obvious in terms of the electrons, it becomes apparent in the bosonic basis. This feature can be spoiled by the non-linearity of the electron dispersion, as discussed in Sec.~\ref{sec:band_curvature}.

Here, we have considered a bosonic number state. Coherent bosonic states can be prepared using a coherent drive. Alternatively, coherent states can be prepared by tunneling an electron into or out of a specific point in the wire, for example by using a scanning tunneling microscope tip. This follows from the bosonization identity $c_z^\dagger \sim e^{i \phi(z)}$ relating the electron creation operator and the local bosonic field $\phi(z)$, which can in turn be expressed in terms of the mode operators $b_k^\dagger$, $b_k^{\phantom\dagger}$~\cite{stone_bosonization_1994}.

\subsection{Electron-hole pairs}

\label{sec:phpair}

\begin{figure}
\begin{center}
\includegraphics[width = 7cm]{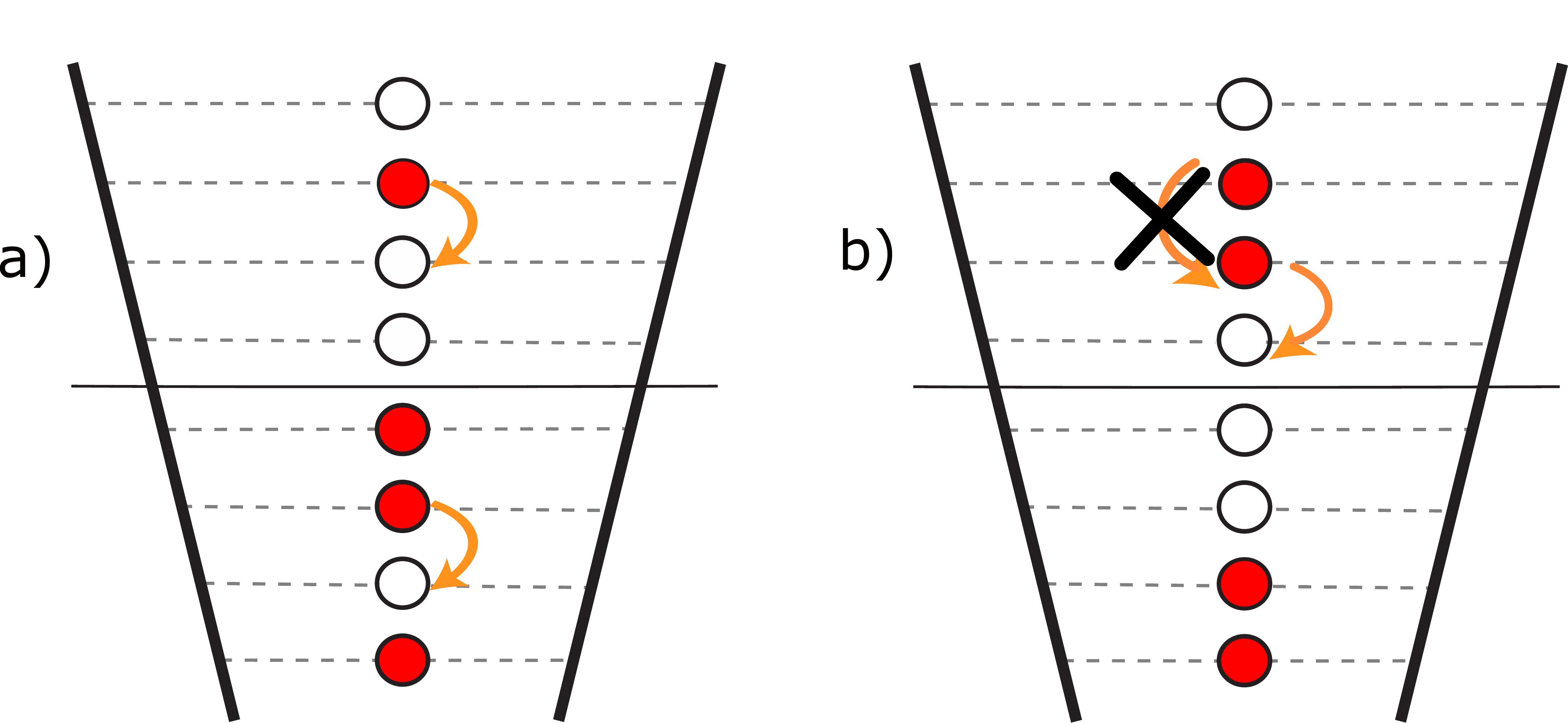}
\caption{(a) Transition $|| 1 \rangle \! \rangle \to || 2 \rangle \! \rangle$ for the electron-hole pair, with $||1 \rangle \! \rangle = |3,3\rangle_{eh}$ and $||2 \rangle \! \rangle = ( |3,2\rangle_{eh} + |2,3 \rangle_{eh} )/\sqrt{2}$. (b) The decay of a two-electron state with $R = 1$. The decay of the more energetic electron is Pauli blocked, as indicated by the ``X''.}
\label{fig:decay}
\end{center}
\end{figure}

Any excited state of an electron gas involves the creation of electron-hole pairs. We consider the cascade initiated by the creation of a single electron-hole pair: $|| 1 \rangle \! \rangle = | N_e , N_h \rangle_{eh}$, where
$| i , j \rangle_{eh} = c_{i}^\dagger c_{-j+1}^{\phantom\dagger} | G\rangle$ 
has an electron and hole in the single-particle levels $i$ and $-j+1$, respectively. The state $|3,3\rangle_{eh}$ is depicted Fig.~\ref{fig:decay}a. (Initializing the system in this state is subtle and it cannot be generated by simply allowing the ground state to absorb a single photon of energy $5 \hbar \omega_1$.) The final state in the cascade is the ground state $|G\rangle$.

For $m < N_e,N_h$, neither the electron nor the hole has reached the top of the Fermi sea at $j \sim 0$ where it would be Pauli blocked. By applying Eq.~(\ref{eq:demotion}) repeatedly, we obtain the wave function
\begin{equation}
|| m \rangle \! \rangle = \frac{1}{\sqrt{\mathcal{N}_m}} \sum_{k = 0}^{m-1} \binom{m-1}{k} | N_e-k,N_h-m+k+1 \rangle_{eh},
\label{eq:ehwf}
\end{equation}
where 
\begin{equation}
\mathcal{N}_m = \binom{2m-2}{m-1}.
\end{equation}
For example, the third state in the cascade is
\begin{eqnarray}
|| 3 \rangle \! \rangle &=& \frac{1}{\sqrt{6}} \bigg( | N_e-2, N_h \rangle_{eh} + 2 | N_e-1,N_h-1 \rangle_{eh} \nonumber \\ 
& & + | N_e, N_h-2 \rangle_{eh} \bigg).
\end{eqnarray}

That $|| m \rangle \! \rangle$ is normalized follows from the combinatoric identity known as Vandermonde's convolution~\cite{graham_concrete_1994}. From the relation
\begin{equation}
\mathcal{D}_1 || m \rangle \! \rangle = \sqrt{\frac{\mathcal{N}_{m+1}}{\mathcal{N}_{m}}} || m + 1 \rangle\rangle
\end{equation}
and Eq.~(\ref{eq:gamma_n}), we find that the rate at which the state $|| m \rangle \! \rangle$ decays is 
\begin{equation}
\Gamma_{m} = \left( 4 - \frac{2}{m} \right) \gamma_1.
\label{eq:phgamma}
\end{equation}
The initial rate $2 \gamma_1$ for $m = 1$ reflects the independence of the electron and hole (the rate of decay of a single electron or hole is $\gamma_1$). As the decay proceeds, the rate increases due to superradiance: photon emission generates entanglement between the electron and hole, as is clear from the wave function (\ref{eq:ehwf}). For $1 \ll m < N_e,N_h$, the rate (\ref{eq:phgamma}) approaches $(1+1)^2 \gamma_1 = 4 \gamma_1$. This indicates that photon emission is fully coherent: the quantum mechanical amplitudes (rather than the rates) add.

The generalization to multiple electron-hole pairs is straightforward. As long as none of the electrons or holes are Pauli blocked, either at the top of the Fermi sea or by each other, the many-body wave function is obtained by replacing the binomial coefficients that appear in Eq.~(\ref{eq:ehwf}) by multinomial coefficients. For $P$ electron-hole pairs, the initial decay rate is $2 P \gamma_1$. The rate approaches $(2P)^2 \gamma_1$ after many steps (but before any electron or hole becomes Pauli blocked).

\begin{figure}
\begin{center}
\includegraphics[width = 8cm]{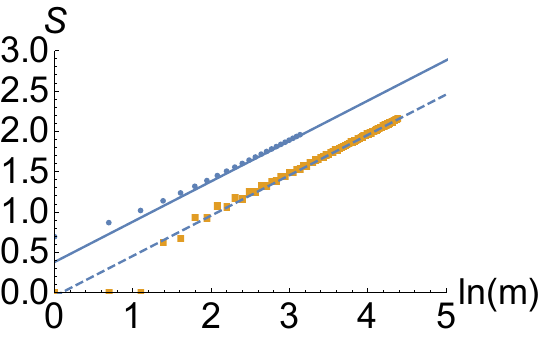}
\caption{Entanglement entropy between an electron and hole (circles) and two electrons (squares) versus $\ln m$. The solid and dotted lines indicate the asymptotic behavior of the entanglement entropy for these cases, respectively. The solid line is given by Eq.~(\ref{eq:e-entropy-approx}).}
\label{fig:entropy}
\end{center}
\end{figure}

The emission rate (\ref{eq:phgamma}) is increasing in $m$ due to the growing entanglement between the electron and hole. This is an example of momentum-space entanglement entropy~\cite{flynn_momentum_2023}. The entanglement entropy of the electron-hole pair is
\begin{equation}
S = - \sum_k p_k \ln p_k,
\label{eq:e-entropy}
\end{equation}
where $p_k = \binom {m-1}{k}^2 / \mathcal{N}_m$. As shown in Fig.~\ref{fig:entropy}, the entanglement entropy grows as the cascade proceeds. For large $m$, a Gaussian approximation can be used to estimate $p_k$, which gives
\begin{equation}
S = \frac{1}{2} \ln m + c,
\label{eq:e-entropy-approx}
\end{equation}
where $c = \ln \frac{\sqrt{\pi}}{2} + \frac{1}{2} \approx 0.38$. The entanglement entropy is a measure of the effective number of electron-hole pair states in the superposition (\ref{eq:ehwf}). The factor of $1/2$ appearing in Eq.~(\ref{eq:e-entropy-approx}) is natural given that the wave function has a width $\sqrt{m}$ in the large $m$ limit.

\begin{figure}
\begin{center}
\includegraphics[width = 7cm]{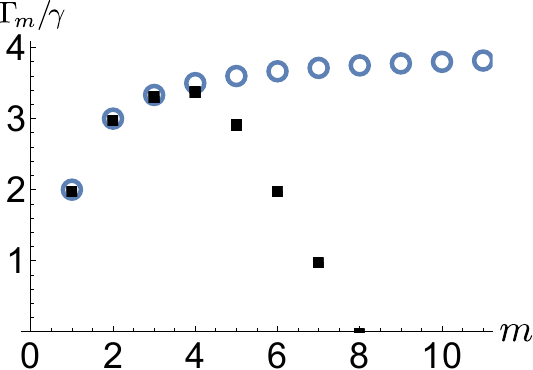}
\caption{The decay rates $\Gamma_m/\gamma_1$ (black squares)  of the states in the cascade with initial state $||1 \rangle \! \rangle = |3,3\rangle_{eh}$. For $m \leq 3$, the rate is consistent with Eq.~(\ref{eq:phgamma}) (open circles). Once $m > 3$, there is an amplitude for the electron or hole to reach the Fermi surface. This suppresses the rate below that predicted by Eq.~(\ref{eq:phgamma}).}
\label{fig:particle-hole}
\end{center}
\end{figure}

For $m > N_e,N_h$, there is amplitude for the electron or the hole to reach the top of the Fermi sea and thus become Pauli blocked. At this point in the cascade, the wave function is no longer given by Eq.~(\ref{eq:ehwf}) and the decay rates will be less than those in Eq.~(\ref{eq:phgamma}). In Fig.~\ref{fig:particle-hole}, the rates $\Gamma_m$ for the cascade $N_e = N_h = 3$ are shown. For $m \leq 3$, the rates conform to Eq.~(\ref{eq:phgamma}). However, for $m > 3$, the rates fall to zero as the system returns to the ground state.

We now consider a cascade that maps directly to the Dicke model. The initial state is formed by promoting each of the $\bar{N}$ most energetic electrons in the ground state by an energy $\bar{N} \hbar \omega_1$, i.e., $|| 1 \rangle \! \rangle$ is $|\bar{N} \bar{N} \bar{N} ... \rangle_f$. The final state is the ground state. In this case, we take the resonant cavity mode to have an energy $\bar{N} \hbar \omega_1$, i.e., $n = \bar{N}$. 

Due to the Pauli exclusion principle, each electron can only decay once and thus can be represented by a single Dicke spin (1/2). In order to demonstrate this correspondence, we will calculate the rates and states for $\bar{N} = 3$. Consider the initial state $|| 1 \rangle \! \rangle = |333\rangle_f$. Subsequent states in the cascade are found by repeatedly applying the lowering operator $b_3$. The state $|333\rangle_f$ can be expressed as a superposition of twelve bosonic basis states. The second state in the cascade, $||2\rangle\rangle$ is given by $b_3|333\rangle_f$ and takes the form
\begin{eqnarray}
||2 \rangle \! \rangle = &-& \frac{1}{2 \sqrt{15}} | 0 \rangle_a+ \frac{1}{2 \sqrt{3}} | 0 \rangle_b - \frac{1}{\sqrt{6}} | 0 \rangle_d \\ &-& \frac{1}{\sqrt{15}} | 0 \rangle_e + \sqrt{\frac{2}{3}} | 2 \rangle_0 ,
\label{eq:state2}
\nonumber
\end{eqnarray}
where the state $|2\rangle_0$ has $2$ quanta in the $m=3$ mode while the other modes (not shown) are in their ground state. The states $| 0 \rangle_x$ with $x = a, b, d, e$ have $0$ quanta in the $m = 3$ bosonic mode but have other modes that are excited. We find
\begin{eqnarray}
\langle \! \langle 2 || b_3 || 1 \rangle \! \rangle &=& 1, \\
\langle \! \langle 3 || b_3 || 2 \rangle \! \rangle &=& \frac{2}{\sqrt{3}}, \\
\langle \! \langle 4 || b_3 || 3  \rangle \! \rangle &=& 1.
\end{eqnarray}
The decay rates (\ref{eq:gamma_n}) are proportional to the square of these amplitudes and follow the ratio $3 : 4 : 3$. 

This can be compared to the Dicke model with $\bar{N} = 3$ spins. The states in the Dicke model are characterized by the magnetic quantum number $M = J, J-1,...,-J$, which is the $z$-component of the total angular momentum $J = N/2$ of the spins. The rate of decay of the state with magnetic quantum number $M$ is~\cite{gross_superradiance_1982}
\begin{equation}
\Gamma_M \propto \left(J + M \right) \left( J - M + 1 \right).
\label{eq:dicke}
\end{equation}
Indeed, we find that the rates of decay (\ref{eq:dicke}) also conform to the ratio $3 : 4: 3$ for $J = \frac{3}{2}$ with $M = \frac{3}{2}, \frac{1}{2}, -\frac{1}{2}$.

Preparing the system in the state $|333\rangle_f$ is likely to be experimentally challenging. A general strategy for preparing the system in complicated states is to excite the electron gas with photons. This yields a bosonic basis state. Then, the desired state can be post-selected. In this case, $|333\rangle_f$ can be prepared by exciting the gas (in its ground state) by two $4 \hbar \omega_1$ photons and one $1 \hbar \omega_1$ photon. This is the state $|1002\rangle_b$. Writing this state as a superposition of electronic states $|...\rangle_f$ reveals how $|333\rangle_f$ can be obtained through post-selection. In particular, a conductance measurement is performed that determines whether the single-electron states $j = 2$ and $j = 3$ are occupied. If both states are found to be occupied, then the measurement will have collapsed the wave function to $|333\rangle_f$. This procedure is based on the detailed form of the unitary connecting the bosonic and electron bases, see  Appendix~\ref{sec:unitary}.

\subsection{Pair of Electrons}

We now consider a cascade in which two electrons are initially excited. This case is more complex due to the fact that the lower energy electron can Pauli block the higher energy one. We consider two electrons far above a filled Fermi sea that are initially separated by $R$ energy levels. This initial state is denoted $|| 1 \rangle \! \rangle = | N_{e} , N_e - R \rangle_{ee}$, where $N_e \gg R$ and the state $| j , k \rangle_{ee} = c_{j}^\dagger c_{k}^{\dagger} | G \rangle$ consists of electrons in the single particle states $j$ and $k$ above a filled Fermi sea.

For $m < R$, there have not been a sufficient number of decays for the top electron to be Pauli blocked. Thus, the form of the wave function and the rates are the same as those of the electron and hole for $m < N$. However, once $m > R$, Pauli blocking comes into play, as illustrated in Fig.~\ref{fig:decay}b. The electronic wave function is
\begin{equation}
|| m \rangle \! \rangle = \frac{1}{\sqrt{\mathcal{N}'_m}}   \sum_{k = 0}^{m-1}  \eta_k | N_e - k, N_e - R - m + k + 1 \rangle_{ee},
\label{eq:eewf}
\end{equation}
where
\begin{equation}
\eta_k = \left[ \binom{m-1}{k} - \binom{m-1}{k-R} \right]_+, 
\end{equation}
and $[x]_+ = x$ for $x > 0$ and $0$ otherwise. We use the convention that the second term in brackets is zero for $k < r$. The normalization constant $\mathcal{N}'_m$ does not have a simple form. The first term in brackets is the same that appears in the wave function (\ref{eq:ehwf}). The second term accounts for Pauli blocking. The quantity $\eta_k$ appears in a closely related combinatoric problem: it is the number of possible paths in a ``truncated'' version of Pascal's triangle, which has a vertical absorbing wall~\cite{donnelly_counting_2018}. In that context, the first term in $\eta_k$ is the total number of paths that arrive at a point, while the second is the number of paths that also hit the absorbing wall. 

\begin{figure}
\begin{center}
\includegraphics[width = 7cm]{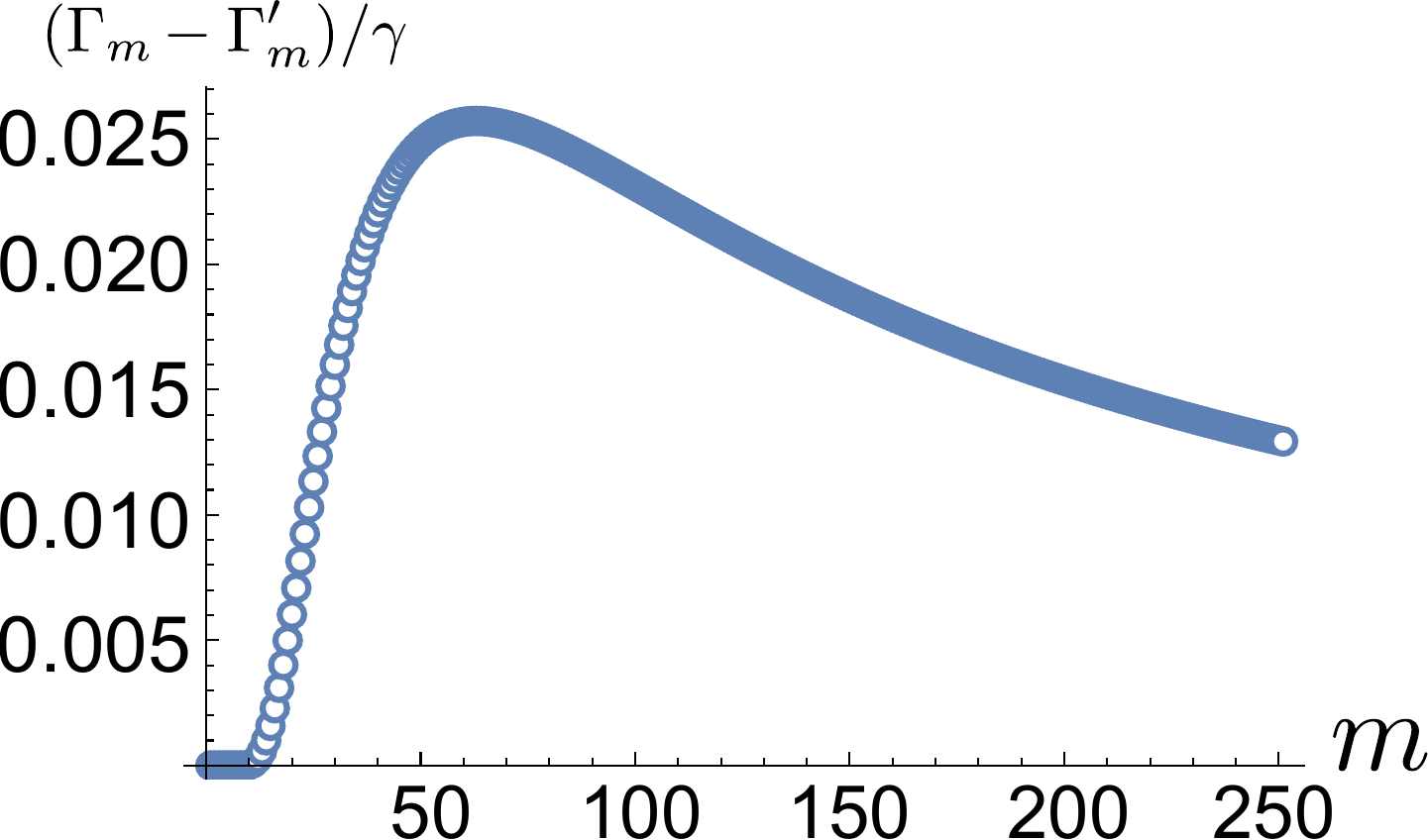}
\caption{The difference in decay rates $(\Gamma_m - \Gamma_m')/\gamma_1$ for the $m^{\textrm{th}}$ step in a cascade involving an electron-hole pair and two electrons, as a function $m$. This difference is a direct measure of the effects of Pauli blocking.}
\label{fig:ee}
\end{center}
\end{figure}

The effects of Pauli blocking are evident in the behavior of $\Gamma_m - \Gamma_m'$, where $\Gamma_m$ is given by Eq.~(\ref{eq:phgamma}). We consider an initial state $|N,N-R\rangle_{ee}$, where $N \gg 1$ and $R = 10$. As shown in Fig.~\ref{fig:ee}, $\Gamma_m = \Gamma_m'$ for $m < R = 10$. For $m > R$, $\Gamma_m' < \Gamma_m$, as is expected. Interestingly, the difference $\Gamma_m-\Gamma_m'$ exhibits non-monotonic behavior in $m$, peaking at $m \approx 50$. Figure~\ref{fig:ee} is suggestive of the fact that $\Gamma_m - \Gamma_m'$ tends to zero for $m$ large (but still $<N$). Below, we argue that this is indeed the case.

The entanglement entropy for the case of two electrons can be defined in a manner analogous to the case of an electron and hole. In Fig.~\ref{fig:entropy}, we have plotted the entanglement entropy between two electrons for $R = 1$. For small $m$, there is a conspicuous alternating pattern in the entanglement entropy $S$. This parity effect arises because of the form of $\mathcal{D}_1$. In particular, if the state $||m\rangle \! \rangle$ contains a Pauli blocked electron, $||m+1 \rangle \! \rangle$ cannot. The asymptotic behavior of $S$ for large $m$ can be estimated from $\eta_k$ using a Gaussian approximation for the binomial coefficients. We find that the entanglement entropy again takes the asymptotic form given by Eq.~(\ref{eq:e-entropy-approx}) where now $c \approx -0.045$. This constant is less than $c$ for the electron-hole case due to Pauli blocking. In the large $m$ limit, the Gaussian approximation gives $D_1 || m \rangle \! \rangle \simeq  2 || m \rangle \! \rangle$, with corrections that vanish as $m \to \infty$. Together with Eq.~(\ref{eq:gamma_n}), this shows that $\Gamma_m' \to 4 \gamma_1$. Since $\Gamma_m \to 4 \gamma_1$ as well (see Eq.~(\ref{eq:phgamma})), we find that $\Gamma_m - \Gamma_m'$ indeed tends to zero for $m \to \infty$, as claimed above.

\section{Phase Coherence}

\label{sec:band_curvature}

Phase coherence is crucial for superradiance. Here, we discuss several important ways in which phase coherence is maintained in an electron gas and the ways it can be lost. A feature of the Dicke model is that the system remains in the spin multiplet with the largest total angular momentum $J$~\cite{gross_superradiance_1982}. This requires that all the spins couple to the field with the same strength. For the electron gas, phase coherence is protected in several ways. First, the operator $\mathcal{D}_n$ remains coherent because of the linearity of the electron dispersion. This guarantees that every operator in the sum $\mathcal{D}_n$ has the same dynamical phase. Phase coherence is also maintained by the form of $\mathcal{M}_{jk}$. The matrix elements are functions of $j-k$ but depend only weakly on $j+k$ (see Appendix~\ref{sec:wfs}). Thus, the phase coherence arises from the universal properties of the 1D electron gas. Unlike the original Dicke model~\cite{garraway_dicke_2011}, these features do not rely on the fine-tuning of the Hamiltonian or the spatial homogeneity of the cavity mode.

There are a number of ways in which phase coherence can be lost. For concreteness, we explore the effect of the electron mass on the dispersion $\omega_j$. Consider a massive dispersion $\omega_j$ in Eq.~(\ref{eq:H0}) given by
\begin{equation}
\omega_j= \frac{ \pi v_F}{L} j + \frac{\pi^2 \hbar }{2 M_e L^2} j^2,
\label{eq:parabola}
\end{equation}
where $M_e$ is the electron mass and $\mathcal{N}$ is the total number of electrons. The integer $j$ indexes the single-particle electron states, see Fig.~\ref{fig:states}a. We consider a cascade with the initial state $||1\rangle\! \rangle$ given by $|\Psi(t) \rangle = |20\rangle_b$. Its time dependence can be written as
\begin{equation}
|\Psi(t)\rangle = \frac{1}{\sqrt{2}} \left( |2 \rangle_f + e^{i \Omega t} | 11 \rangle_f \right),
\label{eq:t}
\end{equation}
where the offset energy $\Omega = \omega_2 - 2 \omega_1 \propto 1/M_e$ arises from the non-linearity of the dispersion. Here, we work in the regime that $\Omega$ is much less than the width of the cavity mode. Writing Eq.~(\ref{eq:t}) in the bosonic basis, we have
\begin{equation}
|\Psi(t)\rangle = \frac{1}{2} \left[ \left(1 + e^{i \Omega t} \right) |2 0 \rangle_b + \left( 1 - e^{i \Omega t} \right) | 0 1 \rangle_b \right].
\end{equation}
For decay rates $\gamma_1 \gg \Omega$, the state $|\Psi(t)\rangle$ decays before it evolves significantly in time. In this regime, the decay rate is $2 \gamma_1$, i.e., the same as $|2 0\rangle_b$. For $\Omega \gg \gamma_1$, the decay rate is $\gamma_1$ and, as expected, we find that the loss of phase coherence tends to reduce the rate of photon emission.

This example shows that for non-linear dispersions, the bosonic states are no longer eigenstates of the electron Hamiltonian. The significance of the linear dispersion is that it exhibits particle-hole symmetry. The particle-hole symmetry operator, which we denote by $\mathcal{P}$, exchanges electron creation operators and hole creation operators. For example, consider the action of the particle-hole symmetry operator on the state formed by promoting the topmost electron in the ground state by $ 2 \hbar \omega_1$, i.e., $\mathcal{P} |2 \rangle_f$. The resultant state can be formed by \emph{demoting} the bottom-most \emph{hole} by the same energy, $|11\rangle_f$. Thus,
\begin{equation}
\mathcal{P} | 2 \rangle_f = | 1 1 \rangle_f.
\end{equation}
Since $\mathcal{P}^2 = 1$, we also have that $ \mathcal{P} | 1 1 \rangle_f = | 2 \rangle_f$~\footnote{This definition of particle-hole symmetry differs from the standard many-body definition taken in Ref.~\cite{chiu_classification_2016}, for example. There, $\mathcal{P}^2 = -1$ because $\mathcal{P}$ is anti-unitary~\cite{chiu_classification_2016}.}. It then follows that $|2 0 \rangle_b$, given by
\begin{equation}
|2 0 \rangle_b = \frac{1}{\sqrt{2}} \left( |2 \rangle_f + | 11 \rangle_f \right),
\end{equation}
is an eigenstate of $\mathcal{P}$ with eigenvalue $+1$. This is not an accident---it turns out that every bosonic basis state $|l_1 l_2 ...\rangle_b$ is an eigenstate of $\mathcal{P}$ with an eigenvalue $(-1)^{\ell_2 + \ell_4 + ...}$. For a linear spectrum, the particle-hole operator $\mathcal{P}$ commutes with the Hamiltonian and so the states $|20\rangle_b$ and
\begin{equation}
|0 1 \rangle_b = \frac{1}{\sqrt{2}} \left( |2 \rangle_f - | 11 \rangle_f \right)
\end{equation}
are guaranteed to be degenerate.

\begin{figure}
\begin{center}
\scalebox{1}[-1]{\includegraphics[width = 7cm]{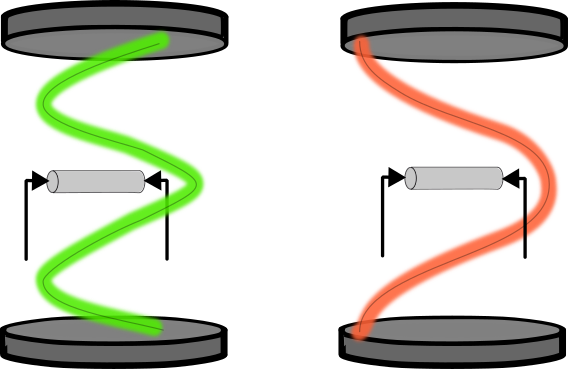}}
\caption{Non-trivial fluorescence of an electron gas arising from a non-linear single-electron dispersion. (left) The electron system is excited by the absorption of a $3\hbar \omega_1$-photon. After a time $\sim 1/\Omega$, the system can decay by emitting a photon with energy $\hbar \omega_1$  (right).}
\label{fig:QW-nonlin}
\end{center}
\end{figure}

An interesting consequence of a massive dispersion is that it alters the fluorescence properties of the electron gas---its fluorescence is no longer trivial, as defined in Sec.~\ref{sec:cascades}.  Consider the situation depicted in Fig.~\ref{fig:QW-nonlin} in which an electron gas with finite $M_e$ is initially prepared in the state $|001\rangle_b$ through the absorption of a $3 \hbar \omega_1$-photon. The state $|001\rangle_b$ can be expressed in terms of the electron basis states, which are now the energy eigenvalues of the system (see Eq.~(\ref{eq:001b})). After a time $\sim 1/\Omega$, the time evolution of the wave function leads to a reasonable probability of being in the $|300\rangle_b$ state. If the system is also coupled to a mode of frequency $\omega_1$, it can then emit photons into this mode, thus giving rise to non-trivial fluorescence.

We briefly remark that even for strong electron-electron interactions, the system's excitations can still be described by a gas of bosons~\cite{stone_bosonization_1994}. This approach to handling interactions is well-known in the context of the Luttinger liquid model. For strong interactions, there is no longer a simple transformation between the electron excitations and the bosons. While this complicates state preparation, the bosonic basis still accounts for the radiative properties of the liquid. It is worth mentioning that certain interactions can give rise to anharmonic couplings between the bosons, but such coupling is weak~\cite{degottardi_thermal_2019}.

\section{Summary and Outlook}

\label{sec:conclusion}

In this work, we have investigated the radiative properties of a one-dimensional electron gas in the quantum coherent regime. Generically, the cascades studied exhibit Dicke-like superradiance that is often frustrated by the effects of Pauli blocking. Our results demonstrate that the finite-sized 1D electron gas offers a new and experimentally accessible platform to study Dicke model physics and its generalization to a many-body system. An interesting aspect of this work is that it demonstrates that a many-body wave function can protect phase coherence. Here, the coherence of a superposition of states is maintained by the equal energy spacing of the single-particle states. This prevents decoherence that would arise from a dynamic phase. This work also suggests that superradiance is a more general feature of many-body systems than previously appreciated.

The 1D electron gas offers a surprisingly rich test bed to explore light-matter coupling in the quantum coherent regime, offering insights into the interplay between Fermi statistics, phase coherence, and quantum entanglement. This is particularly significant given the central role that quantum entanglement plays in various applications, such as quantum computing and quantum transduction. An important extension of this work would be to consider the regime in which quantum information encoded in the electron gas is transferred to the photonic mode.

The authors gratefully acknowledge productive conversations with Mike Stone and Smitha Vishveshwara.

\appendix{}

\section{Single-particle wave functions and the many-body dipole operator}

\label{sec:wfs}

The quantum wire, shown in Fig.~\ref{fig:QW}, lies along the $z$-axis with $-L/2 \leq z \leq L/2$. For particle-in-a-box boundary conditions, $\psi_j(\pm L/2) = 0$ with $\mathcal{N}$ even, the single-electron wave functions are
\begin{equation}
\psi_j(z)=\begin{cases}
          \sqrt{\frac{2}{L}} \sin \left(\frac{\pi (j+\mathcal{N})z}{L}\right) \quad &\text{for} \, j \, \text{even}, \\
          \sqrt{\frac{2}{L}} \cos \left(\frac{\pi (j+\mathcal{N}) z}{L}\right) \quad &\text{for } \, j \, \text{odd}. \\
     \end{cases}
\label{eq:wfs}
\end{equation}
For states $j$ and $k$ with the same parity, $\mathcal{M}_{jk}$ vanishes. For states with opposite parity, we have
\begin{equation}
\mathcal{M}_{jk} = \frac{2eL}{\pi^2 (j-k)^2}
\label{eq:matrix_element}
\end{equation}
for $\mathcal{N} \gg 1$ and  $j,k \ll \mathcal{N}$. Corrections go as $\sim 1/\mathcal{N}$.

\section{Representation Theory of $S_N$}

\label{sec:representation}

The symmetric group $S_N$ is the group of permutations of $N$ objects~\cite{papantonopoulou_algebra_2002}. The order of $S_N$ is the number of ways of arranging $N$ distinguishable objects, i.e., $N!$. For example, $|S_3| = 6$. The elements of $S_3$ include the identity, denoted by $(1)(2)(3)$. This is the trivial permutation, which leaves the order of objects unchanged. Three elements of the group involve swapping two objects: $(12)(3)$, $(13)(2)$, and $(1)(23)$. The cycle notation $(12)(3)$ indicates the swapping of the first and second objects. Finally, there are two 3-cycles: $(123)$ and $(321)$. 

The elements of a group can be partitioned into sets known as conjugacy classes. A conjugacy class contains group elements of the same `type'. For example, the elements $(12)(3)$, $(13)(2)$ and $(1)(23)$, which swap exactly two elements, comprise one conjugacy class. For $S_N$, each conjugacy class contains those elements with the same cycle structure.

A representation of $S_N$ is a set of elements obeying the group algebra. Of particular interest are the so-called irreducible matrix representations, which are the most efficient encoding of the group algebra using matrices~\cite{georgi_lie_1999}. Remarkably, there is a one-to-one correspondence between the irreducible representations of a group and its conjugacy classes. A key tool in representation theory is the character table. This table gives the trace of matrices representing the various group elements. As will be seen in Appendix~\ref{sec:unitary}, the character table of $S_N$ is related to the unitary transformation between the fermionic and bosonic bases for the 1D electron gas.

Table I is the character table for the group $S_2$. Each column of a character table corresponds to a conjugacy class of the group. For $S_N$, these conjugacy classes are specified by the number of cycles of a given length. For example, the first column corresponding to the conjugacy class $(2,0)$ contains the elements with two 1-cycles. There is only one group element in this class, namely the identity $(1)(2)$. The conjugacy class $(0,1)$ contains the element with one 2-cycle, i.e., $(12)$.

Each row of a character table corresponds to an irreducible representation (irrep) of the group. These irreps can be represented by Young diagrams, which consist of $N$ boxes. The various shapes represent these irreps~\cite{georgi_lie_1999}. In the trivial representation, each group element is represented by the number $1$. This mapping trivially satisfies the group algebra. The second row in Table I corresponds to the alternating representation in which each group element is represented by the sign of the corresponding permutation, either $\pm 1$. In general, the first column of a character table gives the dimension of the irrep, i.e., the size of the matrices. This is because the first column corresponds to the identity element, and the trace of the identity matrix is equal to its dimension.

\begin{center}
\begin{table}
\ytableausetup{smalltableaux}
\begin{tabular}{|c||c|r|}
  \hline
  \hline
   & $(2,0)$ & $(0,1)$  \\
  \hline
                      &     &         \\
  \ydiagram{2}        & $1$ & $1$     \, \,   \\  [5ex]
  \ydiagram{1,1}      & $1$ & $-1$  \, \,   \\  [5ex]
  \hline
\end{tabular}
\vspace{0.2 in}
\caption{Character table of the symmetric group $S_2$.}
\end{table}
\end{center}

The group $S_2$ only contains one-dimensional representations. This stems from the fact that $S_2$ is abelian. All non-abelian groups necessarily contain at least one representation with a dimension $> 1$ since simple numbers always commute. The group $S_3$ is non-Abelian and has a character table given in Table II. Note that the second irrep has a dimension of 2.   

\begin{center}
\begin{table}
\ytableausetup{smalltableaux}
\begin{tabular}{|c||c|c|r|}
  \hline
  \hline
   & $(3,0,0)$ & $(1,1,0)$ & $(0,0,1)$  \\
  \hline
                      &     &      &     \\
  \ydiagram{3}        & $1$ & $1$  & $1$   \, \, \,  \\  [5ex]
  \ydiagram{2,1}      & $2$ & $0$  & $-1$  \, \, \,  \\  [5ex]
  \ydiagram{1,1,1}    & $1$ & $-1$  & $1$  \, \, \,   \\ [5ex]
  \hline
\end{tabular}
\vspace{0.2 in}
\caption{Character table of the symmetric group $S_3$.}
\end{table}
\end{center}

\section{Explicit Mapping from Fermionic to Bosonic Bases}

\label{sec:unitary}

The unitary transformation between the fermionic and bosonic bases is, up to normalization factors, given by the characters of the symmetric group. As discussed in Appendix~\ref{sec:representation}, the rows and columns of a character table correspond to the irreps and conjugacy classes of a group, respectively. 

There is a one-to-one correspondence between the conjugacy class $(l_1,l_2,l_3,...)$ of $S_N$ and the bosonic basis states $|l_1 l_2...\rangle_b$, where $N = 1 \ell_1 + 2 \ell_2 + ...$ . Similarly, there is a one-to-correspondence between the irreps of $S_N$ and the electron basis states $|\lambda_1 \lambda_2 \lambda_3...\rangle_f$, where the integers $\lambda_1 \geq \lambda_2 \geq \lambda_3 ... \geq 0$ sum to $N$. As discussed in Appendix~\ref{sec:representation}, the irreps can be represented by Young diagrams. Thus, the Young diagrams can also represent electron states. The $i^{th}$ row of the Young diagram contains $\lambda_i$ boxes. For example, the diagram 
\begin{center}
\ydiagram{3}
\end{center}
is a representation of the state $|3\rangle_f$, while the state $|1 1 1\rangle_f$ is represented by
\begin{center}
\ydiagram{1,1,1}
\end{center}
The requirement that the series $\lambda_1, \lambda_2,...$ is not increasing enforces Pauli exclusion principle. This imposes the rule that no row of a Young diagram can have more boxes than the row above it.

The transformation from the bosonic basis to the fermionic one is given by
\begin{equation}
| \lambda_1 \lambda_2 \lambda_3 ... \rangle_f
 = \sum_{C} \frac{1}{\sqrt{m_C}}\chi^R_C |l_1 l_2...\rangle_b,
 \label{eq:unitary_relation}
\end{equation}
where $\chi^R_C$ is the character corresponding to irrep $R$ of conjugacy class $C$ and
\begin{equation}
m_C = 1^{l_1} 2^{l_2} ... l_1 ! l_2 ! ... l_n!.
\end{equation}
The integer $N!/m_C$ is the number of group elements in the conjugacy class $C$. The fact that Eq.~(\ref{eq:unitary_relation}) defines a unitary transformation follows directly from the orthogonality theorem of finite group representation theory~\cite{georgi_lie_1999}.

For example, using Eq.~(\ref{eq:unitary_relation}), we find that $|0 1 \rangle_b = b_2^\dagger |G \rangle$ can be expressed as
\begin{equation}
| 0 1 \rangle_b = \frac{1}{\sqrt{2}} \left( |2 \rangle_f - | 11 \rangle_f \right),
\label{eq:bos-fermi-ex}
\end{equation}
This identity can also be derived by acting the operator relation~(\ref{eq:bosid}) with $n=2$
\begin{equation}
b_2^\dagger = \frac{1}{\sqrt{2}} \sum_{m} c_{m+2}^\dagger c_m^{\phantom\dagger}
\end{equation}
on the ground state. Note that the minus sign in Eq.~(\ref{eq:bos-fermi-ex}) is required by the anticommutation relations obeyed by fermionic operators~\cite{fetter_quantum_2003}. 

As a second example, we diagonalize the operator $\mathcal{D}_1^\dagger \mathcal{D}_1$ in the $3 \hbar \omega_1$ subspace, as discussed in Sec.~\ref{sec:bosonization} of the main text. The matrix given by Eq.~(\ref{eq:matrix}) is diagonalized
\begin{equation}
U^\dagger \left( \mathcal{D}_1^\dagger \mathcal{D}_1 \right) U = \left(
  \begin{array}{rrr}
    3 & 0 & 0 \\
    0 & 1 & 0 \\
    0 & 0 & 0 \\
  \end{array}
\right),
\label{eq:eigen}
\end{equation}
where the unitary matrix 
\begin{equation}
U = \frac{1}{\sqrt{6}} \left(
  \begin{array}{rrr}
    1 & \sqrt{3} & \sqrt{2} \\
    2 &    0     & -\sqrt{2} \\
    1 &  -\sqrt{3} & \sqrt{2} \\
  \end{array}
\right).
\label{eq:U}
\end{equation}
This unitary matrix converts the bosonic basis states ${|300\rangle_b, |110\rangle_b, |001\rangle_b}$ to the electron basis states ${| 3 \rangle_f, |21\rangle_f, | 111\rangle_f}$.  Note that the pattern of signs in $U$ matches those found in Table II. This is a special case of the transformation that appears on the right-hand side of Eq.~(\ref{eq:unitary_relation}). The coefficients in Eq.~(\ref{eq:3f}) for $|3\rangle_f$ are given by the first row of $U$. The inverse transformation is given by $U^\dagger$ since $U$ is unitary. For example,
\begin{equation}
|001\rangle_b = \frac{1}{\sqrt{3}} | 3 \rangle_f - \frac{1}{\sqrt{3}} |21\rangle_f + \frac{1}{\sqrt{3}} |111\rangle_f.
\label{eq:001b}
\end{equation}
These coefficients are given by the last column of $U$. 

We briefly discuss the particle-hole symmetry operator $\mathcal{P}$. As discussed above, each fermionic state is represented by a Young diagram. The action of the particle-hole operator is to reflect this Young diagram along its main diagonal. The resultant diagram is said to be the conjugate of the first. For example, the diagrams corresponding to the states $|3 \rangle_f$ and $|111\rangle_f$, 
\begin{center}
\ydiagram{3} \quad and \quad \ydiagram{1,1,1} \, ,
\end{center}
respectively, are conjugate to one another. Thus, the two corresponding fermionic states are particle-hole conjugates of each other and thus cannot be eigenstates of $\mathcal{P}$. But its clear then that the state $(|3\rangle_f - |111\rangle_f)/\sqrt{2}$, which is the bosonic basis state $|110\rangle_b$, is an eigenstate of $\mathcal{P}$ with eigenvalue $-1$.

In fact, every bosonic basis state is an eigenstate of $\mathcal{P}$. In particular, $|\ell_1 \ell_2 ... \rangle_b$ is an eigenstate of $\mathcal{P}$ with
\begin{equation}
\mathcal{P} | \ell_1 \ell_2 \ell_3 ... \rangle_b = (-1)^{\sum_i l_{2i}} | \ell_1 \ell_2 \ell_3 ... \rangle_b.
\end{equation}
The eigenvalue of the bosonic basis state is therefore determined by the parity of the number of bosons in the modes with even $k$. Because the number of even cycles in a permutation controls whether the sign of the permutation is positive or negative, the eigenvalue of $\mathcal{P}$ turns out to be the sign of the permutations in the corresponding conjugacy class.

\bibliography{main} 

\end{document}